\documentclass[reprint, superscriptaddress, longbibliography, floatfix, twocolumn, nofootinbib, amsmath, amssymb,
 aps, showkeys]{revtex4-2}
\usepackage{mathtools}
\usepackage{nicefrac}
\usepackage{braket}
\usepackage{siunitx}
\usepackage{hyperref}
\usepackage{gensymb}
\usepackage{xcolor}
\usepackage{graphicx}
\usepackage{dcolumn}
\usepackage{bm}
\usepackage{dsfont}
\usepackage{booktabs}

\usepackage{overpic}

\makeatletter
\newcommand*{\rom}[1]{\expandafter\@slowromancap\romannumeral #1@}
\makeatother

\newcommand{\ra}[1]{\renewcommand{\arraystretch}{#1}}

\begin{document}

\preprint{APS/123-QED}

\title{
Indistinguishability of photonic qubits emitted from trapped $^{40}$Ca$^+$ ions \\ via pulsed excitation} 

\author{Pascal Baumgart}
\affiliation{Fachrichtung Physik, Universit\"at des Saarlandes, 66123 Saarbr\"ucken, Germany}
\affiliation{Zentrum für Quantentechnologien (QuTe), Universit\"at des Saarlandes, 66123 Saarbr\"ucken, Germany}

\author{Max Bergerhoff}
\affiliation{Fachrichtung Physik, Universit\"at des Saarlandes, 66123 Saarbr\"ucken, Germany}
\affiliation{Zentrum für Quantentechnologien (QuTe), Universit\"at des Saarlandes, 66123 Saarbr\"ucken, Germany}

\author{Jonas Meiers}
\affiliation{Fachrichtung Physik, Universit\"at des Saarlandes, 66123 Saarbr\"ucken, Germany}
\affiliation{Zentrum für Quantentechnologien (QuTe), Universit\"at des Saarlandes, 66123 Saarbr\"ucken, Germany}

\author{Stephan Kucera}
\thanks{Current address: Luxembourg Institute of Science and Technology (LIST), 4362 Belveaux, Luxembourg}
\affiliation{Fachrichtung Physik, Universit\"at des Saarlandes, 66123 Saarbr\"ucken, Germany}

\author{J\"urgen Eschner}
\email{juergen.eschner@physik.uni-saarland.de}
\affiliation{Fachrichtung Physik, Universit\"at des Saarlandes, 66123 Saarbr\"ucken, Germany}
\affiliation{Zentrum für Quantentechnologien (QuTe), Universit\"at des Saarlandes, 66123 Saarbr\"ucken, Germany}

\date{\today}

\begin{abstract}	
We investigate the indistinguishability of Raman photons generated from two trapped $^{40}$Ca$^+$ ions using few-nanosecond excitation pulses. We elucidate how spontaneous scattering back to the initial state affects Hong-Ou-Mandel interference. We identify the mean number of back-decays as a measurable single-emitter quantity that correlates with achievable interference visibility of photons from two identical emitters.
\end{abstract}


\maketitle

The ability to generate indistinguishable single photons capable of high-contrast Hong-Ou-Mandel (HOM) interference \cite{Hong_1987} is the keystone for implementing entanglement swapping protocols based on Bell state measurements on two photons from two emitters 
\cite{Simon_2003}. 
Macroscopically separated quantum memories have been successfully entangled using a dual-rail polarization-encoded entanglement swapping scheme 
\cite{Dhara_2023, Zippilli_2008} for various platforms such as 
trapped ions \cite{Moehring_2007} or neutral atoms \cite{vanLeent_2022} emitting into free space, 
ions in cavities \cite{Krutyanskiy_2023}, and color centers in diamond \cite{Bernien_2013}. This method for entangling remote quantum memories can be utilized---employing quantum frequency conversion to telecom wavelengths \cite{Bock_2018}---for the implementation of quantum repeater (QR) schemes \cite{Briegel_1998, Kimble_2008, vanLoock_2020} in quantum networks. Another possible application is 
to connect quantum processing units over shorter length scales for distributed quantum computing \cite{OReilly_2024, Main_2025}. 

One method to generate single photons is laser excitation 
of a Raman transition from a stable ground state to a short-lived 
excited state that decays to a third meta-stable state under the emission of a Raman photon. When the meta-stable state is not 
coupled to other states by additional lasers, the Raman photon will be the final photon in the process, ensuring that it is indeed a single photon. However, the indistinguishability of two of these photons remains influenced by the possibility of spontaneous decay back to the ground state and subsequent re-excitation on the driven transition before the final photon is emitted. It was shown that while the spectrum of the Raman photon is not affected by these back-decays, each additional scattering incoherently adds a time-shifted contribution to its temporal wave-packet, thereby broadening it beyond the Fourier limit \cite{Mueller_2017}. Consequently, the visibility of HOM interference between two such photons is reduced, thereby limiting the fidelity of an entanglement swapping operation.

This problem may be avoided by using excitation pulses that are shorter than the lifetime of the excited state, which for trapped ions or atoms typically lies in the few-nanosecond range. 
High interference visibility has been achieved with sub-nanosecond ultraviolet excitation pulses \cite{Maunz_2007, Kim_2020}. Their generation, however, is technically challenging and requires modulating near-infrared light before it is frequency doubled \cite{Lin_2026}. Additionally, sub-lifetime excitation pulses require high laser powers---scaling with the inverse square of the pulse length---to achieve efficient population transfer to the excited state. A more scalable approach is to use acousto-optic modulators (AOMs). They are capable of cutting out laser pulses with pulse lengths on the order of a few nanoseconds, with a high degree of control over the pulse shape and flexibility regarding the pulse repetition rate, which is often 
not provided by pulsed laser systems. Such high degree of control is desired in order to optimize the excitation pulses with respect to the trade-off between photon generation probability and interference properties under realistic experimental boundary conditions. Furthermore, the flexibility of AOM-based pulse generation---together with quantum frequency conversion---facilitates the interfacing of emitters from different platforms with different timing requirements in a heterogeneous quantum network setting.

In this work, we consider single-photon generation with AOM-controlled excitation pulses of few nanosecond length, where the influence of back-decays on the indistinguishability of the photons starts to become noticeable. Single Raman photons are generated from a single trapped $^{40}$Ca$^+$ ion that is excited on the $\mathrm{S}_{1/2}\to\mathrm{P}_{3/2}$ transition at 393\,nm and decays on the $\mathrm{P}_{3/2}\to\mathrm{D}_{5/2}$ transition at 854\,nm. We measure the mean number of back-decays that occur before the final photon emission, for varying excitation pulse lengths and strengths.  
We also measure the HOM visibility of two photons generated from two co-trapped ions and interfering on a 50:50 beam splitter. We identify the mean number of back-decays as an easily accessible quantity that correlates with the achievable HOM visibility. 
As a practical perspective, we discuss the application of the presented setup for the realization of a QR segment in a QR link according to \cite{vanLoock_2020}.

\section{Setup}\label{sec:setup}

\begin{figure}[h]
    \centering
    \includegraphics[width=0.7\linewidth]{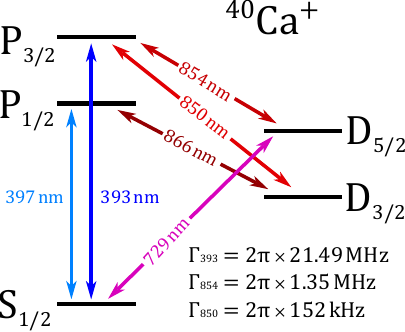}
    \caption{Level scheme of $^{40}$Ca$^+$ with all relevant transition wavelengths and spontaneous decay constants. Laser beams at 397\,nm and 866\,nm are used for Doppler cooling and state detection. A laser at 854\,nm is used for repumping to the ground state. The 393-nm laser is used to generate photons at 854\,nm with parasitic decay at 850\,nm. The laser at 729\,nm is used for electron shelving in order to generate state-selective fluorescence.}
    \label{fig:levelschemeca40}
\end{figure}

The general experimental setup consists of a linear ion trap for  $^{40}$Ca$^+$, stabilized lasers, and single-photon detectors. Various parts of the apparatus have been described in detail in \cite{Bock_2018, Arenskoetter_2024, Bergerhoff_2024}. The relevant transitions, corresponding laser wavelengths and spontaneous decay constants are illustrated in \autoref{fig:levelschemeca40}. 
Short laser pulses at 393\,nm are derived from a continuous-wave laser by a sequence of four AOMs driven by individual, synchronized RF pulses (more details in appendix \ref{appendix:AOMsetup}). The 393-nm laser beam is linearly polarized to excite only $\pi$-transitions and is slightly red-detuned from the $\ket{\mathrm{S}_{1/2},-\frac{1}{2}}\to\ket{\mathrm{P}_{3/2},-\frac{1}{2}}$ transition.

Emitted photons are collected using two in-vacuum high-numerical-aperture laser objectives positioned on either side of the trap along the quantization axis, which is defined by a static magnetic field of approximately 2.85\,G. Photons at 854\,nm are coupled into single-mode fibers and guided to superconducting nanowire single-photon detectors (SNSPDs, \textit{ID Quantique ID281}). Photons at 393\,nm and 397\,nm are coupled into a multi-mode fiber on the opposite side of the trap and detected using a photomultiplier tube (PMT) (\textit{Hamamatsu H7422P-40 SEL}). 

In the first experiment described below in \autoref{sec:expresults}, a single ion is trapped, and detection at 393\,nm is used for acquiring photon correlation data. 
In the second experiment, \autoref{sec:expresults2}, two ions are trapped. The individually collected 854-nm photons are combined on a 50:50 fiber beam splitter (FBS) whose two outputs are guided to the SNSPDs, in order to observe HOM interference. 
The lasers at 397\,nm, 866\,nm, and 729\,nm are additionally used for state discrimination via electron shelving and fluorescence detection, as has already been implemented in \cite{Bergerhoff_2024}. 

\section{Back-decay and photon statistics}\label{sec:expresults}

The mean number of back-decays $\langle N\rangle = \sum_{N=0}^\infty N\cdot P_N$, where $P_N$ is the conditional probability that exactly $N$ back-decays, and thus $N$ emissions of a 393-nm photon, occurred before the emission of a 854-nm Raman photon, is introduced here as a single quantity characterizing the statistics of the back-decays. This number $\langle N\rangle$ and the success probability $P_{854}$ for generating a Raman photon at 854\,nm are measured with a single trapped ion for different excitation pulses, detecting both the emitted 393-nm and 854-nm photons. 
The pulse length is varied between $T_{\mathrm{pulse}}=6.52(6)$\,ns and $T_{\mathrm{pulse}}=56.6(1)$\,ns, where $T_{\mathrm{pulse}}$ is the full width at half maximum (FWHM) of the temporal intensity profile. Only the pulse length is actively varied between measurement runs, but because the AOM setup is generally less efficient for shorter pulses, the laser power at the peak of the pulse also diminishes for shorter pulses. 
In a single cycle of the measurement sequence, the ion is excited with a train of 80 consecutive 393-nm pulses before it is reset to the ground state and laser-cooled. The individual pulses in a train are separated by a repetition period of at least $T_{\mathrm{rep}}\approx 104.25$\,ns; longer repetition periods are used for longer pulses in order to clearly separate them in time. More details are provided in appendix \ref{appendix:AOMsetup}. 
During excitation, photons at 854\,nm and 393\,nm are detected, and their arrival times are recorded with a time-resolution of 625\,ps. We note that every pulse train will generate at most a single 854-nm photon. Additionally, parasitic decay to $\mathrm{D}_{3/2}$ leads to emission of 850-nm photons, limiting the maximally achievable probability to generate an 854-nm photon to $\Gamma_{854} / (\Gamma_{854}+\Gamma_{850})\approx 90$\,\%, where $\Gamma_{854}$ and $\Gamma_{850}$ are the respective spontaneous decay constants.

\begin{figure}[h!]
    \centering
    \includegraphics[width=\linewidth]{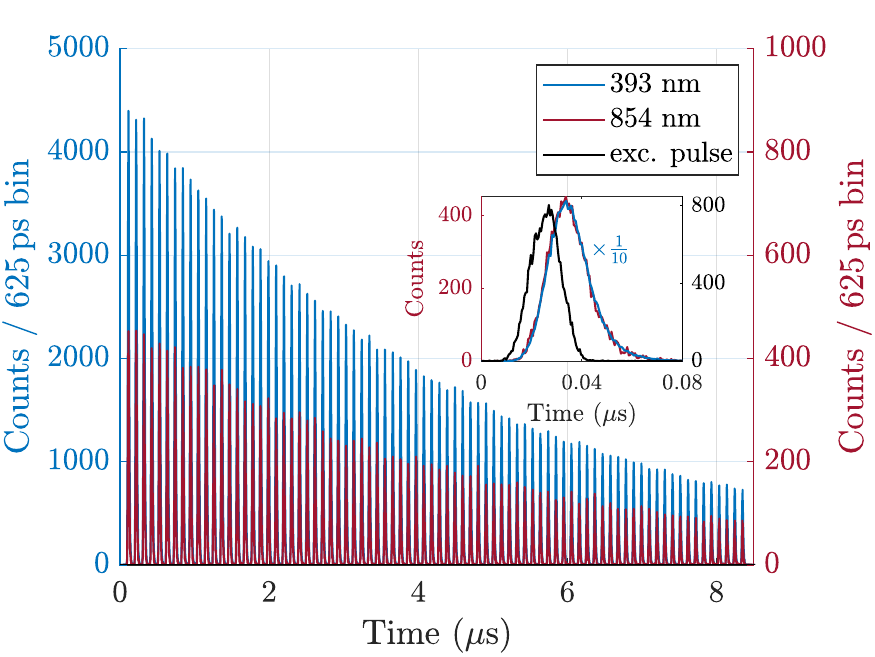}
    \caption{Arrival time distribution of 393-nm (blue) and 854-nm (red) detection events when the ion is excited by a train of 80 pulses of pulse length $T_{\mathrm{pulse}}=15.1(1)$\,ns and spacing $T_{\mathrm{rep}}\approx 104.25$\,ns, reconstructed from $134\times10^6$ cycles of the 
    experimental sequence. Inset: Close-up of the distribution in a time interval corresponding to a single excitation pulse, where the number of 393-nm detection events has been scaled down by a factor of ten to match the heights of the two peaks. Shown in black is the shape of the excitation pulse, reconstructed as outlined in appendix \ref{appendix:AOMsetup}.}
    \label{fig:pulsetrain}
\end{figure}

The reconstructed distribution of arrival times is shown in \autoref{fig:pulsetrain} for the example of $T_{\mathrm{pulse}}=15.1(1)$\,ns. 
One observes that after each of the 80 excitation pulses, the area under the photon wave packet---and thus the ground state population---is reduced by the success probability $P_{\mathrm{Raman}}$ to generate a Raman photon at 850\,nm or 854\,nm. By fitting the decrease of the areas, the success probability $P_{\mathrm{Raman}}$ is extracted. The probability $P_{854}$ of generating a Raman photon specifically at 854\,nm is then given by $P_{854}=\Gamma_{854} / (\Gamma_{854}+\Gamma_{850}) \cdot P_{\mathrm{Raman}}$. 
For the 
parameters of \autoref{fig:pulsetrain}, the success probability 
results as $P_{854}=2.06(1)$\,\%. This value can be extracted from both the 393-nm and 854-nm distributions, and the results always agree within two standard deviations. In the following, the values derived from the 854-nm events are used, due to the lower background of the detectors. 

\begin{figure}[h!]
    \centering
    \includegraphics[width=\linewidth]{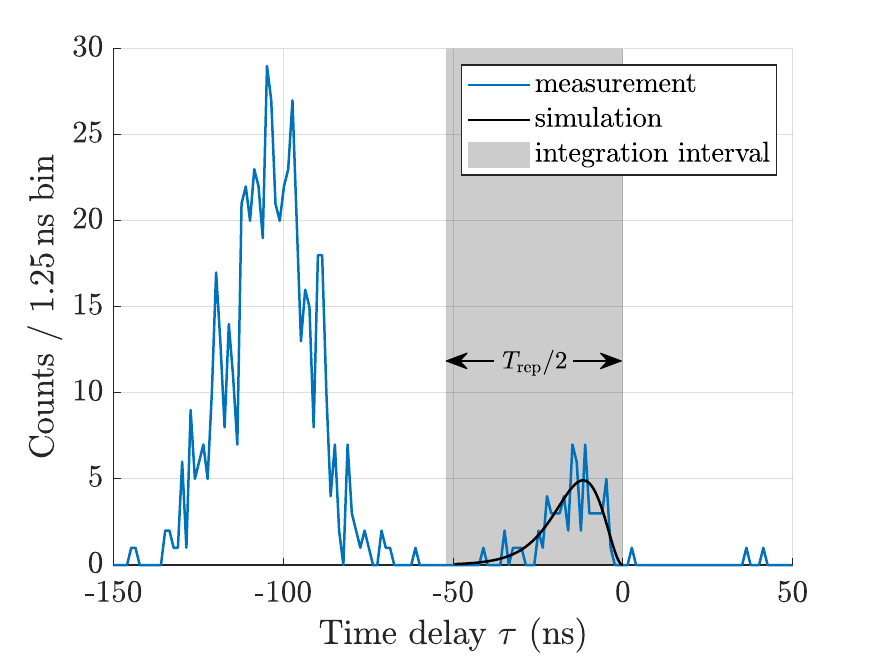}
    \caption{Distribution of detected back-decay photons at 393\,nm for excitation with a pulse length of $T_{\mathrm{pulse}}=15.1(1)$\,ns, reconstructed from $134\times10^6$ cycles of the experimental sequence. The measured cross-correlation between 854-nm and 393-nm photons is shown in blue and compared with a numerical simulation, which is shown as a black line 
    that was scaled to have the same area as the corresponding peak in the measured data. The gray shaded region indicates the integration interval 
    used to determine the mean number of back-decays for this measurement. Its length equals half the pulse repetition period of $T_{\mathrm{rep}}\approx 104.25$\,ns. 
    }
    \label{fig:backdecays}
\end{figure}

The back-decay statistics, in particular $\langle N\rangle$, are extracted from the cross-correlation between the 854-nm and 393-nm detection events, displayed in \autoref{fig:backdecays} for the case   
$T_{\mathrm{pulse}}=15.1(1)$\,ns. 
The figure shows the distribution of 393-nm detection events that precede the detection of an 854-nm photon, as a function of the (negative) time delay. 
As expected, since the emission of the Raman photon leaves the ion in an uncoupled meta-stable state, no events are found for positive time-delays (except for single counts that are explained by background noise). 
The peak around the negative spacing $-T_{\mathrm{rep}}\approx-104.25$\,ns of the pulse train corresponds to 393-nm photons generated by the excitation pulse preceding the one that resulted in the emission of the Raman photon. The relevant information about the back-decay statistics is found in the accumulation of 393-nm detection events just below zero time-delay (shaded region). They will allow the evaluation of the mean number of back-decays $\langle N\rangle$ in the next step. The experimental data agrees well with the theoretical result (black curve), obtained by computing the normally ordered two-time correlation function 
\begin{align}
    \label{eq:g2}
    G_{\mathrm{back-decay}}^{(2)}(t,\tau)=\langle :\hat{n}_1(t-\tau)\hat{n}_2(t):\rangle~.
\end{align}
It describes the number distribution of back-decay photons with the number operator $\hat{n}_1$ emitted with a negative time-delay $\tau$ before a Raman photon with the number operator $\hat{n}_2$ is emitted at time $t$ \cite{Baumgart_2026theo, Fischer_2016}.

Using the measured correlation data as in \autoref{fig:backdecays}, the mean number of back-decays $\langle N\rangle$ is evaluated from the ratio between the total number of 393-nm photons that were emitted just before an 854-nm photon (i.e.\ within the same excitation pulse) and the total number of 854-nm detection events, taking into account the 393-nm detection efficiency $\eta_{393}=3.37(3)\times10^{-3}$ (details in appendix \ref{appendix:393efficiency}). In order to account only for the 393-nm back-decay events that belong to the same excitation pulse that generated the 854-nm photon, the counts within half the repetition period, i.e. for $\tau\in[-T_{\mathrm{rep}}/2,0]$ are considered (shaded region). 

\begin{figure}[h!]
    \centering
    \includegraphics[width=\linewidth]{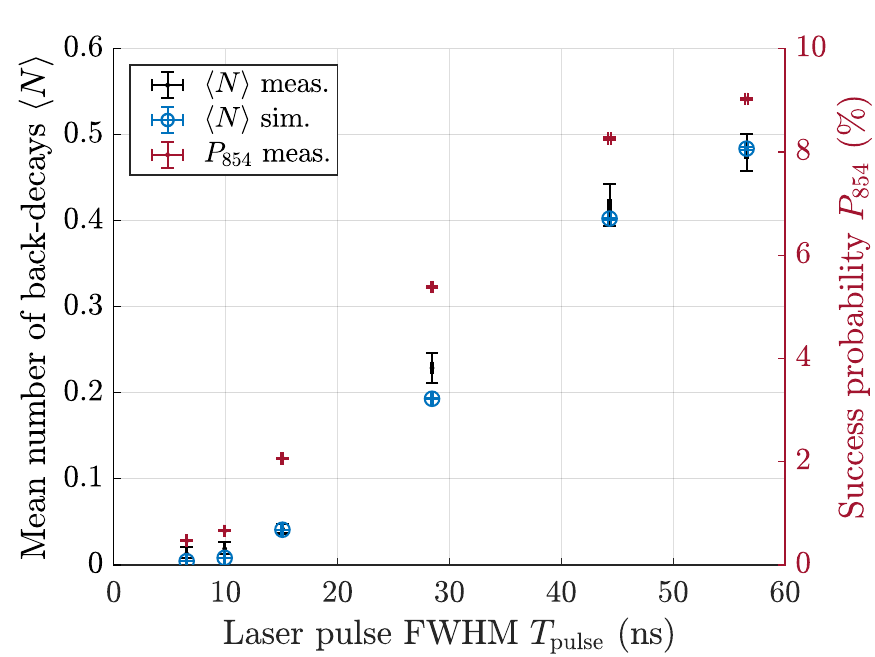}
    \caption{Mean number of back-decays for different excitation pulse lengths $T_{\mathrm{pulse}}$ (FWHM of the intensity profile). The measured data (black) is compared with numerical simulations (blue) that use the measured pulse shape and success probability $P_{854}$ (red) as input parameters. 
    A detailed account of the experimental parameters is given in \autoref{tab:pulseparams} in appendix \ref{appendix:AOMsetup}. 
    }
    \label{fig:nmean}
\end{figure}

The determined values of $\langle N\rangle$ and $P_{854}$ for six different pulse lengths are plotted in \autoref{fig:nmean} and compared to theoretical results for $\langle N\rangle$, calculated from the correlation function \autoref{eq:g2} for the respective 
experimental parameters (see appendix \ref{appendix:AOMsetup}). 

The results show the expected increase of both $\langle N\rangle$ and $P_{854}$
with the length and strength of the excitation pulses.
The good agreement between the measured and calculated values for $\langle N\rangle$ allows us to extend the calculation to other excitation parameters in the next section. 

In the parameter regime covered by the experiment, $\langle N\rangle$ is always less than 0.5, meaning that most of the Raman photons were generated with zero prior back-decays, and that the non-zero value of $\langle N\rangle$ arises mainly from the $N=1$ contribution. This is verified via a quantum-trajectory based theoretical analysis in \cite{Baumgart_2026theo}. As a final remark, while it might appear that the relationship between $\langle N\rangle$ and $P_{854}$ is linear, judging by the fact that the data points seem to follow 
the same curves, this cannot be substantiated without covering a larger parameter space for the excitation pulses. In fact, numerical simulations suggest that there is a more complex relation between them, especially for small $\langle N\rangle$.

\section{Hong-Ou-Mandel interference}\label{sec:expresults2}

\begin{figure}[h!]
    \centering
    \includegraphics[width=0.8\linewidth]{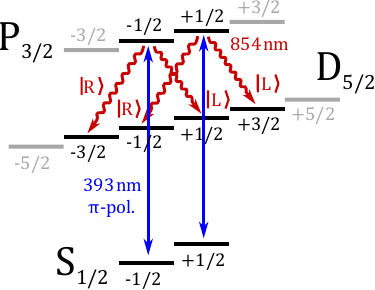}
    \caption{Photon generation scheme relevant for the Hong-Ou-Mandel visibility measurements. The ion is excited with a $\pi$-polarized 393-nm laser pulse resonant to the $\mathrm{S}_{1/2}\to\mathrm{P}_{3/2}$ 
    transition, leading to right- and left-circularly polarized Raman photons at 854\,nm from four different decay channels.}
    \label{fig:schemeHOM}
\end{figure}

With two co-trapped ions and the setup for individual 397-nm and 854-nm photon collection described in \autoref{sec:setup}, the HOM interference is investigated for excitation pulses with lengths between $T_{\mathrm{pulse}}=15.8(1)$\,ns and $T_{\mathrm{pulse}}=55.3(2)$\,ns. The 393-nm excitation was set to $\pi$-polarization, such that 6 different decay paths from the Zeeman substates 
$\ket{\mathrm{P}_{3/2},-\frac{1}{2}}$ and 
$\ket{\mathrm{P}_{3/2},+\frac{1}{2}}$ 
lead to the emission of an 854-nm photon. Those that result in 
a $\pi$-polarized Raman photon need not be considered since they are not emitted along the quantization axis, where photons are collected. 
Thus, as illustrated in \autoref{fig:schemeHOM}, the excitation results in the detection of either right- or left-circularly polarized photons---denoted by the single-photon quantum states $\ket{R}$ and $\ket{L}$---from four different decay paths. The information about the polarization state of a detected photon is inferred from a state readout of both ions, taking advantage of the polarization entanglement between the photons and their respective emitters \cite{Bergerhoff_2024}. We did not differentiate between the two decay paths with the same polarizations. The HOM signal will therefore include photons of equal polarization but different frequency. We verified that this does not significantly influence our results, see appendix \ref{appendix:HOMeval}. 

\begin{figure}[h!]
    \centering
    \includegraphics[width=\linewidth]{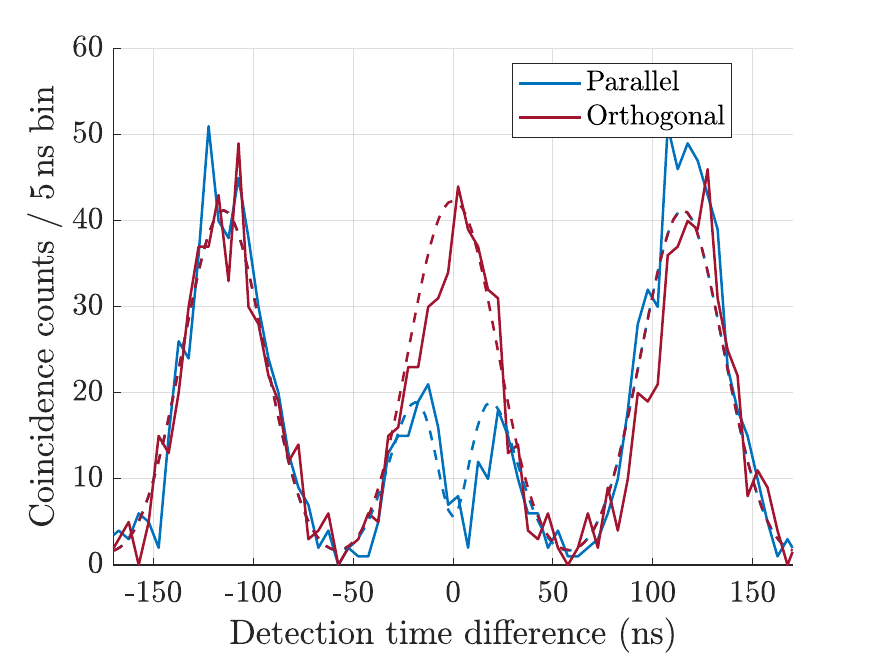}
    \caption{Detection time difference distributions of 854-nm two-photon detection events for ion state readouts projecting onto orthogonal (red) or parallel (blue) polarizations, for an excitation pulse length of $T_{\mathrm{pulse}}=31.7(1)$\,ns. The distribution is reconstructed from $357\times10^7$ cycles of the experimental sequence. The measured data is shown as solid lines and compared with numerical simulations (dashed lines) using experimentally determined parameters for the excitation pulse. }
    \label{fig:dettime}
\end{figure}

As in the previous measurement, the ions are excited by a train of 80 pulses. Here we used five pulse lengths from $T_{\mathrm{pulse}}=15.8(1)$\,ns to $T_{\mathrm{pulse}}=55.3(2)$\,ns, spaced between $T_{\mathrm{rep}}=110$\,ns and $T_{\mathrm{rep}}=140$\,ns (see appendix \ref{appendix:AOMsetup}). The HOM visibility is extracted from the distributions of detection time differences between two registered photons, comparing the results for parallel and orthogonal polarizations (inferred from the ion state readouts). An example is shown in \autoref{fig:dettime} for $T_{\mathrm{pulse}}=31.7(1)$\,ns. 

From these distributions, the HOM visibility is derived as
\begin{align}
    V(T)=1-\frac{C_{\parallel}(T)}{C_\perp(T)} ~,\label{eq:VT}
\end{align}
where $C_\parallel(T)$ and $C_\perp(T)$ are the numbers of two-photon detection events for detection time differences within the coincidence 
window $[-T/2,T/2]$ for photons of parallel and orthogonal polarizations, respectively. 

\begin{figure}[h!]
    \centering
    \includegraphics[width=\linewidth]{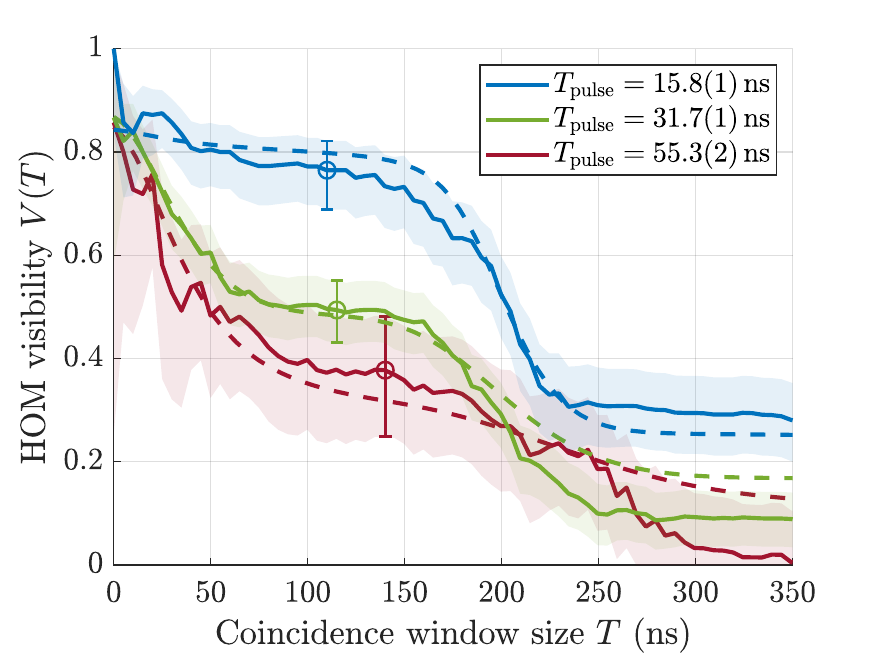}
    \caption{HOM visibility $V(T)$ in dependence of the 
    coincidence window size $T$ for 
    three different pulse lengths $T_{\mathrm{pulse}}=15.8(1)$\,ns (blue), 
    $T_{\mathrm{pulse}}=31.7(1)$\,ns (green) and $T_{\mathrm{pulse}}=55.3(2)$\,ns (red). The measurement results are shown as solid lines, their uncertainties as shaded regions around the data points. Numerical calculations using experimental parameters are shown as dashed lines. Single highlighted data points (circles with uncertainty bars) mark the values $V(T_{\mathrm{rep}})$ evaluated at the pulse repetition period.}
    \label{fig:VHOM}
\end{figure}

The HOM visibility curves, plotted vs. $T$ in \autoref{fig:VHOM} for three out of the five pulse lengths used, show that the visibility approaches its highest value as $T\to0$. At the same time, the 
uncertainty in these values becomes larger due to the decreasing amount of data. 
As the coincidence window size increases, the HOM visibility generally drops but also reaches multiple plateaus, which correspond to the intervals between consecutive excitation pulses in the train. The drop in HOM visibility after each plateau is consistent with the absence of interference for photons generated by different excitation pulses (see appendix \ref{appendix:VHOMres}). The value of interest from the full curve $V(T)$ is the HOM visibility $V(T_{\mathrm{rep}})$ evaluated at the repetition time of the pulses, i.e., for detection time differences within the window $[-T_{\mathrm{rep}}/2, ~T_{\mathrm{rep}}/2]$. This corresponds to taking the full temporal extent of the photon resulting from a single excitation pulse into account. These values are marked as single data points (circles with uncertainty bars) in \autoref{fig:VHOM}.

The measured curves in \autoref{fig:dettime} and \autoref{fig:VHOM} are compared to numerical calculations (dashed lines) using methods from \cite{Woolley_2013, Fischer_2016} with the measured excitation pulse parameters as inputs. For these simulations, we accounted for experimental imperfections, namely background counts, detector efficiency imbalance, decoherence, and an imperfect beam splitter, with their independently determined values. Furthermore, the possibility of a polarization angle mismatch at the FBS that results in the HOM visibility not approaching unity for a small coincidence window size was considered as an additional parameter to fit the numerically calculated HOM visibility curve to the measured data. Details on the inclusion of the experimental imperfections in the simulation are described in appendix \ref{appendix:errormodel}.

\begin{figure}[h!]
    \centering
    \includegraphics[width=\linewidth]{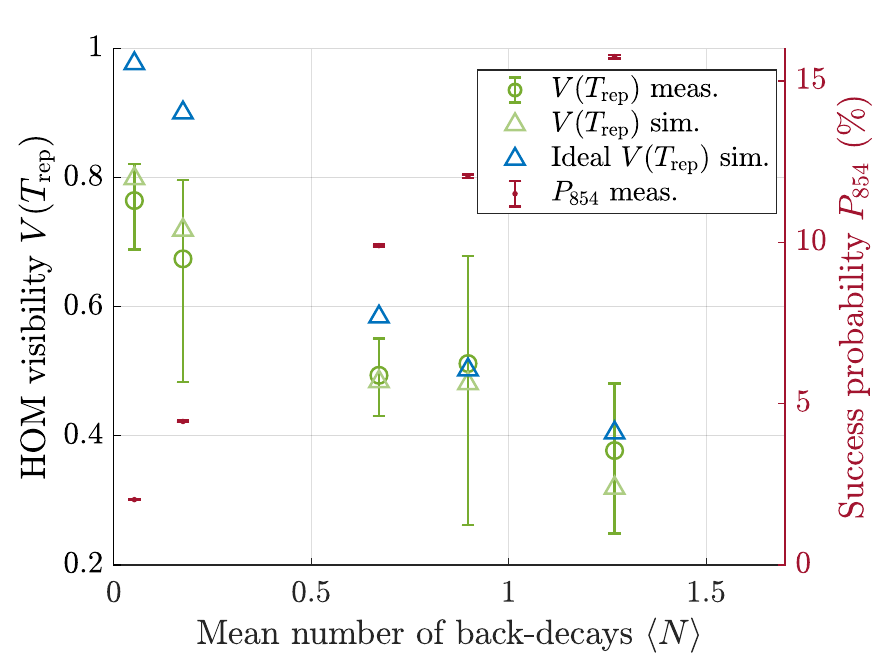}
    \caption{All pairs $(\langle N\rangle,V(T_{\mathrm{rep}}))$ and $(\langle N\rangle,P_{854})$ of values for the mean number 
of back-decays and the HOM visibility or the 
success probability from a series of measurements using excitation pulses of different pulse lengths. The values for the mean number of back-decays were determined numerically using the measured excitation pulse parameters as input. For the HOM visibility, both the measured data (green circles with error bars) and numerically calculated values before (lighter green triangles) and after removing experimental errors (blue triangles) are included. The success probability (red dots with error bars) was measured for each train of excitation pulses using the method described in \autoref{sec:expresults}.}
    \label{fig:HOM_combined}
\end{figure}

The HOM visibility $V(T_{\mathrm{rep}})$ is plotted in \autoref{fig:HOM_combined} vs.\ the mean number of back-decays $\langle N \rangle$, where the latter is numerically calculated for the corresponding excitation pulse. 
Also included is the success probability $P_{854}$, which is measured using the method outlined in \autoref{sec:expresults}. The values for the HOM visibility $V(T_{\mathrm{rep}})$ possess different uncertainties due to unequal data amounts. 
For each data point, two numerically determined values of the HOM visibility, corresponding to imperfect and perfect experimental conditions, are displayed as well. The simulations including imperfections agree well with the measurement. The largest deviations from the ideal HOM visibilities are contributed by the imperfect FBS and by the polarization angle mismatch. The presence of 
background contributes noticeably only for shorter and weaker excitation pulses that result in less signal. Only the polarization angle mismatch was not quantified by an independent measurement, but it is compatible with experience from the laboratory and has 
been identified as a possible source of error by others as well \cite{Meraner_2020}. 
If one assumes that the limitations arise primarily from a miscalibration of the polarization matching at the beam splitter and from imperfections in the beam splitter itself, then one may conclude that it will be feasible to approach the ideal values with reasonable effort. 

The results empirically illustrate how two-photon interference visibility $V(T_{\mathrm{rep}})$ anti-correlates with both
the mean number of back-decays $\langle N\rangle$ and the success probability $P_{854}$. 
This means that the easily accessible property $\langle N\rangle$ of the Raman photons from a single emitter allows insight into the expected HOM visibility in an experiment using two emitters. 
The good agreement of the experimental results with the theoretical description also justifies the use of the numerical methods to investigate the relations between the aforementioned quantities in a broader parameter region and search for an optimum with respect to the trade-off between interference properties and success probability \cite{Baumgart_2026theo}. 

\section{Summary and Discussion}

For single trapped $^{40}$Ca$^+$ ions, the statistics of back-decays during the generation of single Raman photons from a single emitter are shown to be a useful indicator for the indistinguishability of two photons from two identical emitters. This is evidenced by linking the mean number of back-decays $\langle N\rangle$ measured on a single ion with the Hong-Ou-Mandel interference visibility of photons from two ions. 
The results also demonstrate that there is a trade-off between photon indistinguishability and photon generation probability. Numerical simulations are in good agreement with the measurements, demonstrating that the theoretical description may be used in a wider range of parameters and for other ion species. Such a theoretical analysis is currently in preparation \cite{Baumgart_2026theo}. There, among other things, optimization of the excitation pulse parameters under the aforementioned trade-off will be discussed. 

The possibility of generating single photons with high HOM visibility enables the use of the presented setup for realizing quantum repeater operations: indistinguishable single photons entangled with their emitters allow entanglement swapping to two quantum memories. Such a quantum repeater segment has already been demonstrated with two ions in our setup \cite{Baumgart2025_conference}. 
With a second trap setup currently under construction, photonic entanglement swapping can be extended to macroscopically separated memories. Combined with a quantum repeater cell, which we have also realized \cite{Bergerhoff_2024}, two such QR segments form a full QR link \cite{vanLoock_2020}. 
For an implementation over long distances, quantum frequency conversion to the telecom band is necessary, which has been shown to work efficiently with this system \cite{Bock_2018, Arenskoetter_2023}. Furthermore, the flexibility in the timing of the pulses due to the use of AOMs will facilitate the combination of two different platforms---SnV color centers in diamond and trapped ions---into a heterogeneous quantum repeater in future applications. 

Another application is found in quantum key distribution based on atom-photon entanglement with a single ion \cite{Meiers_2024}, where the use of precisely timed short excitation pulses is expected to improve the secret key rate over the one achievable with long excitation pulses.  
Short excitation pulses also enable the generation of time-bin qubits \cite{Ward_2022} or atom–photon entanglement with time-bin encoding \cite{Saha_2025}. In addition, they facilitate the use of polarization-to–time-bin converters \cite{Scalcon_2022}, 
since the temporal separation of the excitation pulses of $\sim 100$\,ns is compatible with interferometer-type polarization-to–time-bin converters needed to interface with systems operating on time-bin qubits \cite{Tchebotareva_2019}.

\section*{Author contributions}
P.\,B. and M.\,B. carried out the experiment and collected the data. P.\,B. and M.\,B. evaluated the experimental data using a script from J.\,M. P.\,B. performed the simulations. P.\,B. and M.\,B. wrote the manuscript with input from all authors.
S.\,K. conceived the nanosecond laser pulse generation setup.
J.\,E. supervised the project.

\begin{acknowledgments}
We acknowledge support from the Federal Ministry of Research, Technology and Space (BMFTR) through projects Q.sync (16KISQ045), QR.X (16KISQ001K), and QR.N (16KIS2180). 
\end{acknowledgments}

\appendix
\section*{Appendix}
\section{Details and characterization of the AOM setup}\label{appendix:AOMsetup}
The setup that cuts out excitation pulses with pulse lengths of a few nanoseconds from a cw laser at 393\,nm consists of 4 free-space acousto-optic modulators (AOMs, \textit{Brimrose TEF-125-100-397}) 
based on tellurium dioxide (TeO$_2$). They are placed in series, alternating between the positive and negative first diffraction order to avoid inducing an overall frequency shift of the light, and are each individually driven using short radio frequency (RF) pulses supplied by a multichannel AWG (\textit{Keysight M3202A}). 
The shape of the RF pulses is sinusoidal with a Gaussian envelope whose width is limited to a minimum value of $\sim 8$\,ns, the inverse of the central frequency of the AOMs of 125\,MHz. 
The beam is focused into each AOM to a focal diameter 
of $\sim 70(5)$\,\unit{\micro\metre}. This further limits the minimal possible pulse length out of a single AOM due to the travel time of the acoustic wave (acoustic velocity $\sim 4200\,\frac{\mathrm{m}}{\mathrm{s}}$) across the diameter of the focus, $\sim12(1)$\,ns. 
With the use of multiple AOMs and time-shifted RF pulses, it is possible to further reduce the optical pulse length at the cost of overall efficiency. The efficiency of the setup also generally decreases as shorter input RF pulse lengths are used, since the response of the AOMs is too slow to reach their maximum diffraction efficiency. 
The shortest pulse length that was achieved with this setup was 
below 5\,ns. It should be noted that the AOMs used in this setup 
are not ideal for their purpose and were only used because they 
were readily available to us. Ideally, AOMs with a large bandwidth 
and a high damage threshold should be used. The TeO$_2$-based 
AOMs turned out to be easily damaged by the required powers 
of 393-nm light. In future applications, the use 
of AOMs based on crystalline quartz will be considered. 

\begin{figure}[h!]
    \centering
    \includegraphics[width=\linewidth]{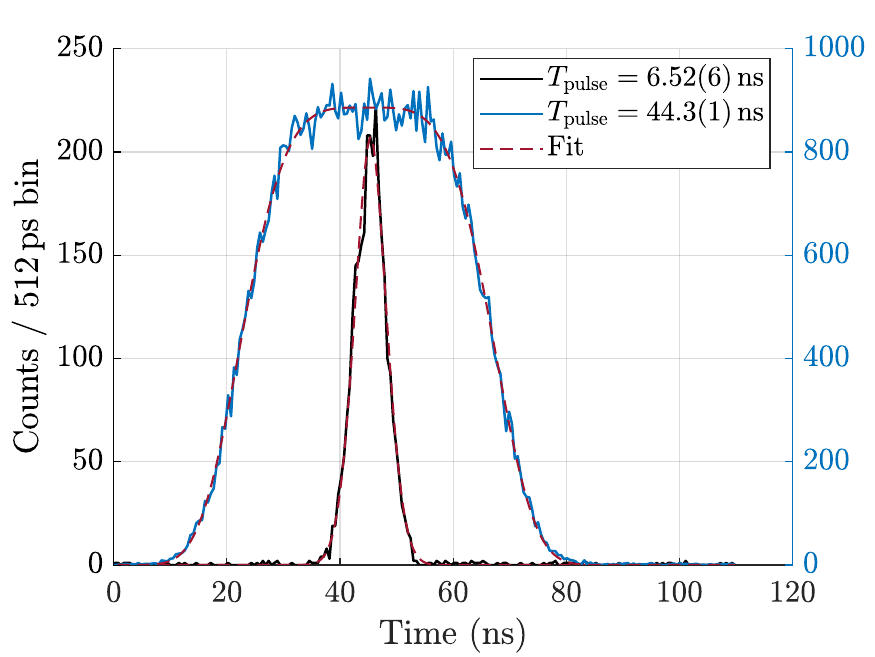}
    \caption{Laser pulses measured with an APD and a photon counter with 512\,ps time resolution. They are shown as histograms of photon detection events after the pulse trigger (solid lines) for two excitation pulses, with $T_{\mathrm{pulse}}=6.52(6)$\,ns (black) and $T_{\mathrm{pulse}}=44.3(1)$\,ns (blue). Measured curves are fitted using the function in \autoref{eq:betafun} (red dashed lines).}
    \label{fig:aompulses}
\end{figure}

In order to perform numerical simulations that match the experimental 
conditions, the shapes of the excitation pulses 
were characterized by connecting the fiber-coupled output of 
the pulse generation setup---after attenuating the 
light---to an avalanche photodiode (\textit{Micro Photon Devices MPD-020-CTF-FC}) and using a time-correlated single-photon counting device (\textit{PicoHarp 300}) 
to correlate its output with a pulse trigger. Two examples 
corresponding to pulse lengths of $T_{\mathrm{pulse}}=6.52(6)$\,ns 
and $T_{\mathrm{pulse}}=44.3(1)$\,ns are shown in \autoref{fig:aompulses} and compared to fitted curves which 
were used to approximate the real pulse shapes in 
numerical simulations. The fit function is a generalized 
Gaussian distribution 
\begin{align}
    f(t)=A\cdot\exp\left(-\left(\frac{\vert t-t_0\vert}{\sqrt{2} \sigma}\right)^\beta\right), \label{eq:betafun}
\end{align}
where $A$ is the amplitude, $t_0$ the position of the 
center of the pulse, $\beta$ a parameter governing the shape of the distribution, and $\sigma$ a measure of the width 
of the pulse. The FWHM is related to the latter via 
$T_{\mathrm{pulse}}=2\sqrt{2}\,(\ln2)^{1/\beta}\,\sigma$. 
This distribution accounts for pulse shapes that deviate from a 
Gaussian distribution by displaying a plateau that 
occurs when the AOMs have reached their maximum diffraction 
efficiency. 

Besides the pulse shape, the Rabi frequency at the peak of the pulse is needed to perform simulations. It is chosen such that the numerically calculated probability $P_{854}$ to generate an 854-nm photon matches the experimentally determined value measured as in \autoref{sec:expresults}. 

The pulse lengths used in the two measurement series in \autoref{sec:expresults} and \autoref{sec:expresults2}, as well as the corresponding pulse shape parameters $\beta$, pulse repetition periods, and numbers of cycles of the experimental sequences (each cycle including 80 excitation attempts), are listed in \autoref{tab:pulseparams}.

\begin{table}[h!]\centering
\ra{1.3}
\begin{tabular}{@{}cccc@{}}\toprule
\multicolumn{4}{c}{$\langle N\rangle$ measurement (\autoref{sec:expresults})} \\ 
$T_{\mathrm{pulse}}$ (ns) & $\beta$ & $T_{\mathrm{rep}}$ (ns) & \#cyc.\,($10^6$)\\
\midrule
$6.52(6)$ & $1.85(4)$ & $104.25$ & $128$\\
$9.91(7)$ & $2.27(5)$ & $104.25$ & $82$\\
$15.08(8)$ & $2.28(4)$ & $104.25$ & $134$\\
$28.45(9)$ & $2.97(4)$ & $114.25$ & $80$\\
$44.3(1)$ & $4.24(6)$ & $159.25$ & $80$\\
$56.6(1)$ & $4.53(6)$ & $169.25$ & $100$\\

\bottomrule
\toprule
\multicolumn{4}{c}{$V(T_{\mathrm{rep}})$ measurement (\autoref{sec:expresults2})} \\ 
$T_{\mathrm{pulse}}$ (ns) & $\beta$ & $T_{\mathrm{rep}}$ (ns) & \#cyc.\,($10^6$) \\
\midrule
$15.8(1)$ & $2.23(5)$ & $110$ & $588$\\
$20.9(1)$ & $2.19(5)$ & $112$ & $338$\\
$31.7(1)$ & $2.63(4)$ & $115$ & $357$\\
$44.2(1)$ & $3.30(4)$ & $140$ & $125$\\
$55.3(2)$ & $3.46(5$) & $140$ & $346$\\
\bottomrule
\end{tabular}
\caption{Pulse lengths $T_{\mathrm{pulse}}$, pulse shape parameters $\beta$ (see \autoref{eq:betafun}), pulse repetition periods 
$T_{\mathrm{rep}}$, and numbers of cycles of the experimental sequence for each series of measurements in \autoref{sec:expresults} and \autoref{sec:expresults2}, respectively.}
\label{tab:pulseparams}
\end{table}

\section{Detection efficiency of 393-nm photons} \label{appendix:393efficiency}
The detection efficiency $\eta_{393}=3.37(3)\times10^{-3}$ of 393-nm photons is measured in two steps. First, the detection efficiency $\eta_{854}=5.18(1)\times10^{-3}$ of 854-nm photons is measured by exciting the ion with a 393-nm pulse whose pulse length is in the microsecond range and thus much longer than the lifetime of the excited $\mathrm{P}_{3/2}$ state. Here, a pulse length of $1.8$\,\unit{\micro\second} was used, long enough to almost completely pump the population from the ground state $\mathrm{S}_{1/2}$ to the two metastable states $\mathrm{D}_{3/2}$ and $\mathrm{D}_{5/2}$. 
From the knowledge of (almost) certainly having created an 854-nm or 850-nm photon with each excitation pulse, the detection efficiency is then determined from the ratio between the number of 
excitation pulses and the number of detected photons. To determine the 393-nm detection efficiency, the ratio between detection events at 393-nm and 854-nm is measured and compared to the branching ratio according to the spontaneous decay constants of the relevant transitions. 

We note that the 854-nm detection efficiency does not factor in when experimentally determining the mean number of back-decays. The number of coincidence detections between a 393-nm and an 854-nm photon is proportional to the product $\eta_{393}\cdot\eta_{854}$ of the two efficiencies, but it is normalized by the total number of detected 854-nm photons, which is itself proportional to the 854-nm detection efficiency, thus eliminating the dependence on $\eta_{854}$. 

\section{HOM visibility evaluation}\label{appendix:HOMeval}
\begin{figure}[h!]
    \centering
    \includegraphics[width=\linewidth]{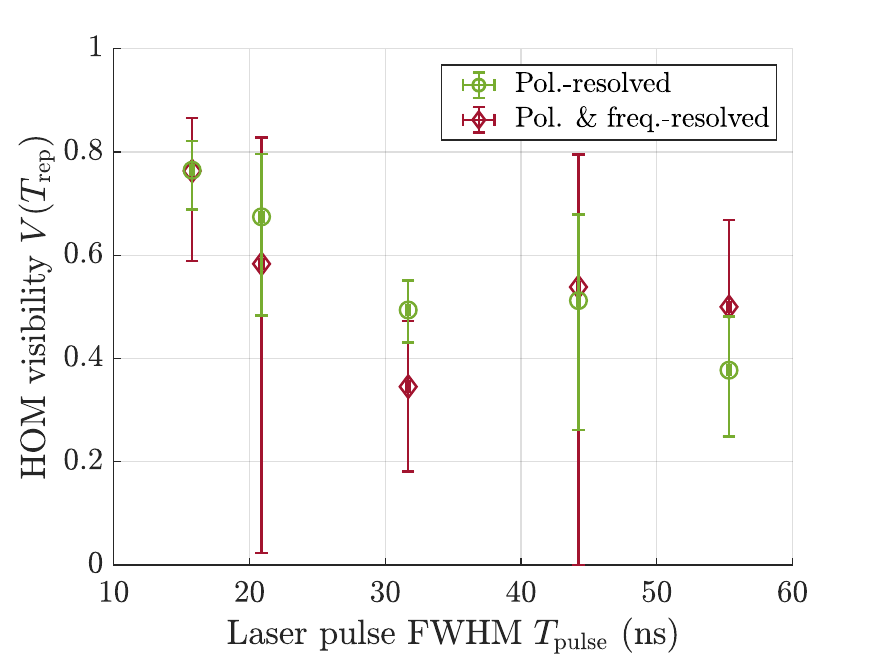}
    \caption{HOM visibility evaluated at the pulse repetition period, $V(T_{\mathrm{rep}})$, for five different excitation pulse lengths between $T_{\mathrm{pulse}}=15.8(1)$\,ns and $T_{\mathrm{pulse}}=55.3(2)$\,ns, using two different methods of evaluation, as described in appendix \ref{appendix:HOMeval}.}
   \label{fig:VHOMeval}
\end{figure}

For the evaluation of the HOM visibility in \autoref{sec:expresults2}, the information about the polarization of detected photons is inferred from a state readout of both ions. When deciding whether a coincidence of two photons corresponds to parallel or orthogonal polarization states, no distinction between the two possible decay paths for each polarization, starting from two different initial Zeeman substates in $\mathrm{S}_{1/2}$ (see \autoref{fig:schemeHOM}), is made. This means that photons with the same polarization may have been emitted via different decay channels and thus have a frequency difference of $\Delta\approx 2\pi \times 3.2$\,MHz (for a static magnetic field of $\sim 2.85$\,G). They are nevertheless entering into $C_\parallel(T)$, the number of coincidence events of photons with state readouts corresponding to parallel polarizations, as introduced in \autoref{eq:VT}. 
This is expected to introduce quantum beats that reduce the HOM visibility for sufficiently large coincidence window size and when the duration of the photon wave packet becomes comparable with the inverse frequency difference $\Delta^{-1}\approx50$\,ns \cite{Legero_2004}. However, within the experimental uncertainties, this effect is not clearly resolved in our case. This is demonstrated by \autoref{fig:VHOMeval}, 
which shows the HOM visibilities evaluated at the pulse repetition period, 
$V(T_{\mathrm{rep}})$, vs.\ the excitation pulse length, extracted from the same data in two different ways. The red data points coincide with those included in \autoref{fig:HOM_combined}; they only differentiate between the polarizations of the photons but not between the decay channels that have the same polarization. For the green data points, in contrast, coincidence events which correspond to photons of the same polarization but generated via different decay channels are not taken into account in the calculation of the HOM visibility. This ensures that coincidence events entering into $C_\parallel(T)$ (assuming ideal readout fidelity and polarization matching) have both the same polarization and frequency. 
The error bars are naturally larger when using the second method. But more crucially, they either extend beyond or at least overlap with the error bars when using the first method. As a result, we decided not to apply a frequency-resolved evaluation of the HOM visibility for the sake of smaller uncertainties.  

\section{Residual HOM visibility}\label{appendix:VHOMres}

\begin{figure}[h!]
    \centering
    \includegraphics[width=\linewidth]{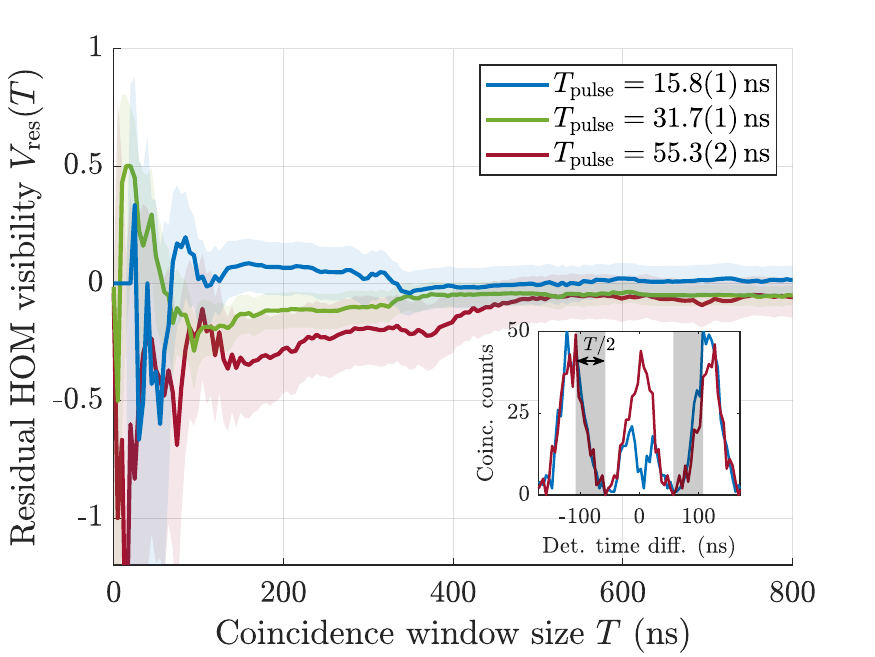}
    \caption{Residual HOM visibility in dependence of the coincidence window size $T$ for three different pulse lengths $T_{\mathrm{pulse}}=15.8(1)$\,ns (blue), 
    $T_{\mathrm{pulse}}=31.7(1)$\,ns (green) and $T_{\mathrm{pulse}}=55.3(2)$\,ns (red). The measurement results are shown as solid lines with the errors as shaded regions around the  data points. Inset: Detection time difference distributions for orthogonal (red) and parallel (blue) polarizations, as shown in \autoref{fig:dettime}, for 
    $T_{\mathrm{pulse}}=31.7(1)$\,ns to illustrate the integration window (gray shaded area) used for the residual HOM visibility. 
    }
    \label{fig:VHOMres}
\end{figure}

In \autoref{sec:expresults2}, the HOM visibility of photons 
generated by trains of excitation pulses is measured for different excitation pulse lengths, with the results shown in \autoref{fig:VHOM}. As the coincidence window size $T$ is increased, the HOM visibility generally decreases, but it also displays multiple plateaus corresponding to the time-intervals between consecutive excitation pulses. The drop in HOM visibility after each plateau is explained by the absence of interference for photons that result from different excitation pulses within a  pulse train. 
In order to verify this, we evaluated specifically the residual HOM visibility $V_\mathrm{res}(T)$
from coincidence data that excludes the window $[-T_{\mathrm{rep}}/2,T_{\mathrm{rep}}/2]$, i.e., using only data from the two windows of size $T/2$ that extend beyond the inner excluded region (see the shaded regions in the inset of \autoref{fig:VHOMres}). 
The result is displayed in \autoref{fig:VHOMres} for the same excitation pulse lengths as in \autoref{fig:VHOM}, $T_{\mathrm{pulse}}=15.8(1)$\,ns, $T_{\mathrm{pulse}}=31.7(1)$\,ns, and $T_{\mathrm{pulse}}=55.3(2)$\,ns. 
Besides initial fluctuations, which are attributed to the small data amounts, the residual HOM visibility converges to zero in all three cases as the coincidence window size is increased. This verifies that there is vanishing interference between photons that are not created by the same excitation pulse.

\section{Computation of the HOM visibility}
\label{appendix:errormodel}

The numerical evaluation of the HOM visibility using the experimental parameters of the excitation pulses is based on calculating normally ordered two-time correlation functions of the electric field operators $E_i(t)=E^{(+)}_i(t)+E_i^{(-)}(t)$, where $E_i^{(+)}(t)=(E_i^{(-)}(t))^\dagger$, for the two output modes $i=c,d$ of a beam splitter, which are related to the two input modes $i=a,b$ by 
\begin{align}
    E_c^{(+)}(t)&=\sqrt{\mathfrak{t}}~
    E_a^{(+)}(t)+i\sqrt{1-\mathfrak{t}}~E_b^{(+)}(t),\label{eq:bstrafo1}\\
    E_d^{(+)}(t)&=
    i\sqrt{1-\mathfrak{t}}~E_a^{(+)}(t)+\sqrt{\mathfrak{t}}~E_b^{(+)}(t), \label{eq:bstrafo2}
\end{align}
where $\mathfrak{t}$ is the power transmission of the beam splitter. 
Note that here, operators are denoted without hats. 
The correlation functions are derived and evaluated numerically by relating them to the dynamics of the ions, modeled by a three-level master equation, using methods described in \cite{Fischer_2016, Woolley_2013}; the exact equations used for the simulations are outlined in \cite{Baumgart_2026theo}. 

As a first experimental imperfection, the transmission of the FBS, which was measured to be $\mathfrak{t}=0.436(11)$ and thus deviates from an ideal 50:50 beam splitter, is considered when applying the beam splitter transformation in \autoref{eq:bstrafo1} and \ref{eq:bstrafo2}.

Losses in fiber transmission from the ions to the inputs of the beam splitter were measured to be almost equal for both paths, so they do not enter. However, the detection efficiencies $\eta_c$, $\eta_d$ (including additional losses) at both outputs were determined to be different by a factor of $\sim 2.14(1)$, which is caused by a combination of fiber connections of different quality at the beam splitter outputs, losses due to additional optics in front of the detectors, and a difference in the quantum efficiencies of the detectors themselves. This is modeled together with background noise by replacing the output fields of the beam splitter with the transformed fields 
\begin{align}
    \tilde{E}_c^{(+)}(t)&= \sqrt{\eta_c}~E_c^{(+)}(t) + B_c^{(+)}(t), \\ 
    \tilde{E}_d^{(+)}(t)&= \sqrt{\eta_d}~E_d^{(+)}(t) + B_d^{(+)}(t). 
\end{align}
Here, $B_{c,d}^{(+)}(t)$ represent the background noise at each output; vacuum fields that enter always where loss occurs have been neglected, as they do not contribute to the correlations. 
The background terms are assumed to be uncorrelated with the fields 
$E_{c,d}^{(+)}(t)$, with each other, and with themselves, i.e. when calculating the normally ordered two-time second-order correlation function with the transformed fields, 
$$\tilde{G}^{(2)}_{\mathrm{HOM}}(t,\tau, \phi)=\langle \tilde{E}_c^{(-)}(t+\tau)\tilde{E}^{(-)}_d(t)\tilde{E}^{(+)}_d(t)\tilde{E}^{(+)}_c(t+\tau)\rangle,$$ 
where $\phi$ is the polarization angle between the two input fields, only contributions with products of background field operators belonging to the same spatial mode that are evaluated at the same time survive and also factorize. The full expression in terms of the original output and background fields is then given by
\begin{align}
    \tilde{G}^{(2)}_{\mathrm{HOM}}&(t,\tau, \phi) = 
    \eta_c\eta_d G^{(2)}_{\mathrm{HOM}}(t,\tau, \phi) \nonumber
    \\ 
    &+ \eta_d\langle E_d^{(-)}(t+\tau)E_d^{(+)}(t+\tau)\rangle \langle B_c^{(-)}(t)B_c^{(+)}(t)\rangle \nonumber \\
    &+\eta_c\langle E_c^{(-)}(t)E_c^{(+)}(t)\rangle \langle B_d^{(-)}(t+\tau)B_d^{(+)}(t+\tau)\rangle \nonumber \\
    &+\langle B_c^{(-)}(t)B_c^{(+)}(t)\rangle\langle B_d^{(-)}(t+\tau)B_d^{(+)}(t+\tau)\rangle, 
\end{align}
where $G^{(2)}_{\mathrm{HOM}}(t,\tau, \phi)$ is the correlation 
function with the original field operators. 
Since the background rates of the detectors are comparable, the background field correlation functions are assumed to be equal for both output ports and constant in time, i.e., 
\begin{align}
    \langle B_{c,d}^{(-)}(t)B_{c,d}^{(+)}(t)\rangle = B_0~.
\end{align}
The numerical value of $B_0$ is determined using the measured signal-to-background ratio of the full single-photon wave packet (inlcuding all 80 excitation pulses) by setting it equal to the calculated ratio 
\begin{align}
    \mathrm{SBR}&=\frac{(\eta_c+\eta_d)\int_0^{t_{\mathrm{end}}}\mathrm{d}t~\langle E_{c}^{(-)}(t)E_{c}^{(+)}(t)\rangle}{2\int_0^{t_{\mathrm{end}}}\mathrm{d}t~\langle B_c^{(-)}B_c^{(+)}(t)\rangle} \nonumber \\
    &=\frac{(\eta_c+\eta_d)\int_0^{t_{\mathrm{end}}}\mathrm{d}t~\langle E_{c}^{(-)}(t)E_{c}^{(+)}(t)\rangle}{2B_0t_{\mathrm{end}}}
\end{align}
after summing up both detectors; $t_{\mathrm{end}}$ is the maximum time considered in the simulation. For both the photon electric field and the background field, it is assumed that the correlation functions for $c$ and $d$ are the same.

Regarding the atomic dynamics, a pure-dephasing term that results in the damping of optical coherence is added to the master equation via the collapse operator $C_{\delta\omega}=\sqrt{2\delta\omega}~\vert e\rangle\langle e\vert$, where $\vert e\rangle$ is the excited state of the ion, and $\delta\omega$ is associated with a finite laser linewidth and was chosen to be $\delta\omega=2\pi \times 50$\,kHz in accordance with typically measured values. 
To account for the fact that a train of multiple excitation pulses is used in the experiments, the numerical calculations are also performed with multiple (but fewer) pulses. However, the RF pulses used to drive the AOMs that generate the optical pulses do not possess a fixed phase relation, and thus, no interference is expected from photons that are generated with different excitation pulses. To account for this, the time evolution (according to equation (7) in \cite{Fischer_2016}) after a photon is measured at time $t$ is restricted to time shifts $\tau$ within the same excitation pulse. It should be noted that this only affects the shape and height of the coincidence peaks corresponding to photons from different excitation pulses (leftmost and rightmost peaks in \autoref{fig:dettime}). Most importantly, it does not change the value of the HOM visibility $V(T_{\mathrm{rep}})$ (see \autoref{sec:expresults2}) corresponding to the full temporal extent of a photon resulting from a single excitation pulse. 

As a final source of limited indistinguishability, an imperfect polarization overlap at the beam splitter is considered by introducing a polarization angle mismatch $\delta\phi$, such that the final calculated HOM visibility is given by
\begin{align}
    V(T)=1-\frac{\int_0^\infty\mathrm{d}t\int_{-T/2}^{T/2}\mathrm{d}\tau~\tilde{G}^{(2)}_{\mathrm{HOM}}(t,\tau,\delta\phi)}{\int_0^\infty\mathrm{d}t\int_{-T/2}^{T/2}\mathrm{d}\tau~\tilde{G}^{(2)}_{\mathrm{HOM}}(t,\tau,\frac{\pi}{2}+\delta\phi)}.
\end{align}
Since this additional parameter could not be measured independently, 
it is determined by adjusting $\delta\phi$ such that the calculated $V(T)$ agrees best with the measured curve. The best agreement is obtained with assumed polarization angle mismatch values of less than $\delta\phi=\pi/8$. 
\bibliography{Bibliography.bib}

@article{Briegel_1998,
  title = {Quantum Repeaters: The Role of Imperfect Local Operations in Quantum Communication},
  author = {Briegel, H.-J. and D\"ur, W. and Cirac, J. I. and Zoller, P.},
  journal = {Phys. Rev. Lett.},
  volume = {81},
  issue = {26},
  pages = {5932--5935},
  numpages = {0},
  year = {1998},
  month = {Dec},
  publisher = {American Physical Society},
  doi = {10.1103/PhysRevLett.81.5932},
  url = {https://link.aps.org/doi/10.1103/PhysRevLett.81.5932}
}

@article{Dhara_2023,
  title = {Entangling quantum memories via heralded photonic Bell measurement},
  author = {Dhara, Prajit and Englund, Dirk and Guha, Saikat},
  journal = {Phys. Rev. Res.},
  volume = {5},
  issue = {3},
  pages = {033149},
  numpages = {19},
  year = {2023},
  month = {Sep},
  publisher = {American Physical Society},
  doi = {10.1103/PhysRevResearch.5.033149},
  url = {https://link.aps.org/doi/10.1103/PhysRevResearch.5.033149}
}

@article{Zippilli_2008,
doi = {10.1088/1367-2630/10/10/103003},
url = {https://doi.org/10.1088/1367-2630/10/10/103003},
year = {2008},
month = {oct},
publisher = {},
volume = {10},
number = {10},
pages = {103003},
author = {Zippilli, Stefano and Olivares-Rentería, Georgina A and Morigi, Giovanna and Schuck, Carsten and Rohde, Felix and Eschner, Jürgen},
title = {Entanglement of distant atoms by projective measurement: the role of detection efficiency},
journal = {New Journal of Physics}
}

@article{Hong_1987,
  title = {Measurement of subpicosecond time intervals between two photons by interference},
  author = {Hong, C. K. and Ou, Z. Y. and Mandel, L.},
  journal = {Phys. Rev. Lett.},
  volume = {59},
  issue = {18},
  pages = {2044--2046},
  numpages = {0},
  year = {1987},
  month = {Nov},
  publisher = {American Physical Society},
  doi = {10.1103/PhysRevLett.59.2044},
  url = {https://link.aps.org/doi/10.1103/PhysRevLett.59.2044}
}

@article{Simon_2003,
  title = {Robust Long-Distance Entanglement and a Loophole-Free Bell Test with Ions and Photons},
  author = {Simon, Christoph and Irvine, William T. M.},
  journal = {Phys. Rev. Lett.},
  volume = {91},
  issue = {11},
  pages = {110405},
  numpages = {4},
  year = {2003},
  month = {Sep},
  publisher = {American Physical Society},
  doi = {10.1103/PhysRevLett.91.110405},
  url = {https://link.aps.org/doi/10.1103/PhysRevLett.91.110405}
}

@article{Kim_2020,
author = {Junki Kim and Junho Jeong and Changhyun Jung and Minjae Lee and Yunjae Park and Dong-il Dan Cho and Taehyun Kim},
journal = {Opt. Express},
number = {26},
pages = {39727--39738},
publisher = {Optica Publishing Group},
title = {Observation of Hong-Ou-Mandel interference with scalable {Y}b$^+$-photon interfaces},
volume = {28},
month = {Dec},
year = {2020},
url = {https://opg.optica.org/oe/abstract.cfm?URI=oe-28-26-39727},
doi = {10.1364/OE.409667},
}

@article{Moehring_2007,
author={Moehring, D. L.
and Maunz, P.
and Olmschenk, S.
and Younge, K. C.
and Matsukevich, D. N.
and Duan, L.-M.
and Monroe, C.},
title={Entanglement of single-atom quantum bits at a distance},
journal={Nature},
year={2007},
month={Sep},
day={01},
volume={449},
number={7158},
pages={68-71},
issn={1476-4687},
doi={10.1038/nature06118},
url={https://doi.org/10.1038/nature06118}
}

@article{Meraner_2020,
  title = {Indistinguishable photons from a trapped-ion quantum network node},
  author = {Meraner, M. and Mazloom, A. and Krutyanskiy, V. and Krcmarsky, V. and Schupp, J. and Fioretto, D. A. and Sekatski, P. and Northup, T. E. and Sangouard, N. and Lanyon, B. P.},
  journal = {Phys. Rev. A},
  volume = {102},
  issue = {5},
  pages = {052614},
  numpages = {14},
  year = {2020},
  month = {Nov},
  publisher = {American Physical Society},
  doi = {10.1103/PhysRevA.102.052614},
  url = {https://link.aps.org/doi/10.1103/PhysRevA.102.052614}
}

@article{Maunz_2007,
author={Maunz, P.
and Moehring, D. L.
and Olmschenk, S.
and Younge, K. C.
and Matsukevich, D. N.
and Monroe, C.},
title={Quantum interference of photon pairs from two remote trapped atomic ions},
journal={Nature Physics},
year={2007},
month={Aug},
day={01},
volume={3},
number={8},
pages={538-541},
issn={1745-2481},
doi={10.1038/nphys644},
url={https://doi.org/10.1038/nphys644}
}

@article{Lin_2026,
    author = {Lin, Jianlong and Cieszynski, Mari and Christopherson, William and Khan, Darman and Li, Lintao and Goldschmidt, Elizabeth and DeMarco, Brian},
    title = {{$^{88}$Sr$^+$} ion trap apparatus for generating 408\,nm photons},
    journal = {Review of Scientific Instruments},
    volume = {97},
    number = {1},
    pages = {013201},
    year = {2026},
    month = {01},
    issn = {0034-6748},
    doi = {10.1063/5.0288146},
    url = {https://doi.org/10.1063/5.0288146},
}

@Article{vanLeent_2022,
author={van Leent, Tim
and Bock, Matthias
and Fertig, Florian
and Garthoff, Robert
and Eppelt, Sebastian
and Zhou, Yiru
and Malik, Pooja
and Seubert, Matthias
and Bauer, Tobias
and Rosenfeld, Wenjamin
and Zhang, Wei
and Becher, Christoph
and Weinfurter, Harald},
title={Entangling single atoms over 33{\thinspace}km telecom fibre},
journal={Nature},
year={2022},
month={Jul},
day={01},
volume={607},
number={7917},
pages={69-73},
issn={1476-4687},
doi={10.1038/s41586-022-04764-4},
url={https://doi.org/10.1038/s41586-022-04764-4}
}

@Article{Bernien_2013,
author={Bernien, H.
and Hensen, B.
and Pfaff, W.
and Koolstra, G.
and Blok, M. S.
and Robledo, L.
and Taminiau, T. H.
and Markham, M.
and Twitchen, D. J.
and Childress, L.
and Hanson, R.},
title={Heralded entanglement between solid-state qubits separated by three metres},
journal={Nature},
year={2013},
month={May},
day={01},
volume={497},
number={7447},
pages={86-90},
issn={1476-4687},
doi={10.1038/nature12016},
url={https://doi.org/10.1038/nature12016}
}

@article{OReilly_2024,
  title = {Fast Photon-Mediated Entanglement of Continuously Cooled Trapped Ions for Quantum Networking},
  author = {O'Reilly, Jameson and Toh, George and Goetting, Isabella and Saha, Sagnik and Shalaev, Mikhail and Carter, Allison L. and Risinger, Andrew and Kalakuntla, Ashish and Li, Tingguang and Verma, Ashrit and Monroe, Christopher},
  journal = {Phys. Rev. Lett.},
  volume = {133},
  issue = {9},
  pages = {090802},
  numpages = {6},
  year = {2024},
  month = {Aug},
  publisher = {American Physical Society},
  doi = {10.1103/PhysRevLett.133.090802},
  url = {https://link.aps.org/doi/10.1103/PhysRevLett.133.090802}
}

@Article{Main_2025,
author={Main, D.
and Drmota, P.
and Nadlinger, D. P.
and Ainley, E. M.
and Agrawal, A.
and Nichol, B. C.
and Srinivas, R.
and Araneda, G.
and Lucas, D. M.},
title={Distributed quantum computing across an optical network link},
journal={Nature},
year={2025},
month={Feb},
day={01},
volume={638},
number={8050},
pages={383-388},
issn={1476-4687},
doi={10.1038/s41586-024-08404-x},
url={https://doi.org/10.1038/s41586-024-08404-x}
}

@article{Mueller_2017,
  title = {Spectral properties of single photons from quantum emitters},
  author = {M\"uller, Philipp and Tentrup, Tristan and Bienert, Marc and Morigi, Giovanna and Eschner, J\"urgen},
  journal = {Phys. Rev. A},
  volume = {96},
  issue = {2},
  pages = {023861},
  numpages = {9},
  year = {2017},
  month = {Aug},
  publisher = {American Physical Society},
  doi = {10.1103/PhysRevA.96.023861},
  url = {https://link.aps.org/doi/10.1103/PhysRevA.96.023861}
}

@article{Bergerhoff_2024,
  title = {Quantum repeater node with free-space coupled trapped ions},
  author = {Bergerhoff, Max and Elshehy, Omar and Kucera, Stephan and Kreis, Matthias and Eschner, J\"urgen},
  journal = {Phys. Rev. A},
  volume = {110},
  issue = {3},
  pages = {032603},
  numpages = {12},
  year = {2024},
  month = {Sep},
  publisher = {American Physical Society},
  doi = {10.1103/PhysRevA.110.032603},
  url = {https://link.aps.org/doi/10.1103/PhysRevA.110.032603}
}

@article{Fischer_2016,
  title = {Dynamical modeling of pulsed two-photon interference},
  author = {Fischer, Kevin A and Müller, Kai and Lagoudakis, Konstantinos G and Vučković, Jelena},
  journal = {New J. Phys.},
  volume = {18},
  pages = {113053},
  year = {2016},
  doi = {10.1088/1367-2630/18/11/113053},
  url = {https://iopscience.iop.org/article/10.1088/1367-2630/18/11/113053/meta}
}

@article{Krutyanskiy_2023,
  title = {Entanglement of Trapped-Ion Qubits Separated by 230 Meters},
  author = {Krutyanskiy, V. and Galli, M. and Krcmarsky, V. and Baier, S. and Fioretto, D. A. and Pu, Y. and Mazloom, A. and Sekatski, P. and Canteri, M. and Teller, M. and Schupp, J. and Bate, J. and Meraner, M. and Sangouard, N. and Lanyon, B. P. and Northup, T. E.},
  journal = {Phys. Rev. Lett.},
  volume = {130},
  issue = {5},
  pages = {050803},
  numpages = {7},
  year = {2023},
  month = {Feb},
  publisher = {American Physical Society},
  doi = {10.1103/PhysRevLett.130.050803},
  url = {https://link.aps.org/doi/10.1103/PhysRevLett.130.050803}
}

@article{Bock_2018,
 title={High-fidelity entanglement between a trapped ion and a telecom photon via quantum frequency conversion},
 volume={9},
 number={1},
 journal={Nature Communications},
 author={Bock, Matthias and Eich, Pascal and Kucera, Stephan and Kreis, Matthias and Lenhard, Andreas and Becher, Christoph and Eschner, Jürgen},
 year={2018},
 url= {https://doi.org/10.1038/s41467-018-04341-2}}

@article{Kimble_2008,
  title = {The quantum internet},
  author = {Kimble, H. J.},
  journal = {Nature},
  volume = {453},
  issue = {7198},
  pages = {1023},
  numpages = {7},
  year = {2008},
  month = {Jun},
  publisher = {Nature},
  doi = {10.1038/nature07127},
  url = {https://doi.org/10.1038/nature07127}
}

@article{vanLoock_2020,
author = {van Loock, Peter and Alt, Wolfgang and Becher, Christoph and Benson, Oliver and Boche, Holger and Deppe, Christian and Eschner, Jürgen and Höfling, Sven and Meschede, Dieter and Michler, Peter and Schmidt, Frank and Weinfurter, Harald},
title = {Extending {Q}uantum {L}inks: {M}odules for {F}iber- and {M}emory-{B}ased {Q}uantum {R}epeaters},
journal = {Advanced Quantum Technologies},
volume = {3},
number = {11},
pages = {1900141},
doi = {https://doi.org/10.1002/qute.201900141},
url = {https://onlinelibrary.wiley.com/doi/abs/10.1002/qute.201900141},
year = {2020}
}

@article{Woolley_2013,
doi = {10.1088/1367-2630/15/10/105025},
url = {https://doi.org/10.1088/1367-2630/15/10/105025},
year = {2013},
month = {oct},
publisher = {IOP Publishing},
volume = {15},
number = {10},
pages = {105025},
author = {Woolley, M J and Lang, C and Eichler, C and Wallraff, A and Blais, A},
title = {Signatures of {Hong–Ou–Mandel} interference at microwave frequencies},
journal = {New Journal of Physics}
}

@inproceedings{Meiers_2024,
author = {Jonas Meiers and Christian Haen and Max Bergerhoff and Stephan Kucera and J\"{u}rgen Eschner},
booktitle = {Quantum 2.0 Conference and Exhibition},
journal = {Quantum 2.0 Conference and Exhibition},
pages = {QTh3A.27},
publisher = {Optica Publishing Group},
title = {Quantum key distribution with atom-photon entanglement for an urban fiber link},
year = {2024},
url = {https://opg.optica.org/abstract.cfm?URI=QUANTUM-2024-QTh3A.27},
doi = {10.1364/QUANTUM.2024.QTh3A.27},
}

@unpublished{Baumgart_2026theo,
  author       = {Pascal Baumgart and Max Bergerhoff and Jürgen Eschner},
  title        = {Indistinguishability of single {Raman} photons from single atoms},
  note         = {Manuscript in preparation},
  year         = {2026}
}

@article{Saha_2025, 
title={High-fidelity remote entanglement of trapped atoms mediated by time-bin photons}, 
volume={16}, 
number={1}, 
journal={Nature Communications}, 
author={Saha, Sagnik and Shalaev, Mikhail and O’Reilly, Jameson and Goetting, Isabella and Toh, George and Kalakuntla, Ashish and Yu, Yichao and Monroe, Christopher}, 
year={2025},
url={https://doi.org/10.1038/s41467-025-57557-4}, 
month={Mar}}

@article{Ward_2022,
doi = {10.1088/1367-2630/aca9ee},
url = {https://doi.org/10.1088/1367-2630/aca9ee},
year = {2022},
month = {dec},
publisher = {IOP Publishing},
volume = {24},
number = {12},
pages = {123028},
author = {Ward, Travers and Keller, Matthias},
title = {Generation of time-bin-encoded photons in an ion-cavity system},
journal = {New Journal of Physics}
}

@article{Tchebotareva_2019,
  title = {Entanglement between a Diamond Spin Qubit and a Photonic Time-Bin Qubit at Telecom Wavelength},
  author = {Tchebotareva, Anna and Hermans, Sophie L. N. and Humphreys, Peter C. and Voigt, Dirk and Harmsma, Peter J. and Cheng, Lun K. and Verlaan, Ad L. and Dijkhuizen, Niels and de Jong, Wim and Dr\'eau, Ana\"{\i}s and Hanson, Ronald},
  journal = {Phys. Rev. Lett.},
  volume = {123},
  issue = {6},
  pages = {063601},
  numpages = {6},
  year = {2019},
  month = {Aug},
  publisher = {American Physical Society},
  doi = {10.1103/PhysRevLett.123.063601},
  url = {https://link.aps.org/doi/10.1103/PhysRevLett.123.063601}
}

@inproceedings{Baumgart2025_conference,
author = {Pascal Baumgart and Max Bergerhoff and Jonas Meiers and J\"{u}rgen Eschner},
booktitle = {Optica Quantum 2.0 Conference and Exhibition},
journal = {Optica Quantum 2.0 Conference and Exhibition},
pages = {QTh4A.3},
publisher = {Optica Publishing Group},
title = {Entanglement swapping with two co-trapped {$^{40}$Ca$^+$} ions for quantum repeater application},
year = {2025},
url = {https://opg.optica.org/abstract.cfm?URI=QUANTUM-2025-QTh4A.3},
doi = {10.1364/QUANTUM.2025.QTh4A.3},
}

@Article{Arenskoetter_2023,
author={Arensk{\"o}tter, Elena
and Bauer, Tobias
and Kucera, Stephan
and Bock, Matthias
and Eschner, J{\"u}rgen
and Becher, Christoph},
title={Telecom quantum photonic interface for a {$^{40}$Ca$^+$} single-ion quantum memory},
journal={npj Quantum Information},
year={2023},
month={Apr},
day={10},
volume={9},
number={1},
pages={34},
issn={2056-6387},
doi={10.1038/s41534-023-00701-z},
url={https://doi.org/10.1038/s41534-023-00701-z}
}

@article{Scalcon_2022,
author = {Scalcon, Davide and Agnesi, Costantino and Avesani, Marco and Calderaro, Luca and Foletto, Giulio and Stanco, Andrea and Vallone, Giuseppe and Villoresi, Paolo},
title = {Cross-Encoded Quantum Key Distribution Exploiting Time-Bin and Polarization States with Qubit-Based Synchronization},
journal = {Advanced Quantum Technologies},
volume = {5},
number = {12},
pages = {2200051},
doi = {https://doi.org/10.1002/qute.202200051},
url = {https://advanced.onlinelibrary.wiley.com/doi/abs/10.1002/qute.202200051},
year = {2022}
}

@article{Arenskoetter_2024,
  title = {{Full Bell-basis measurement of an atom-photon 2-qubit state and its application for quantum networks}},
  author = {Arensk\"otter, Elena and Kucera, Stephan and Elshehy, Omar and Bergerhoff, Max and Kreis, Matthias and Brunel, L\'eandre and Eschner, J\"urgen},
  journal = {Phys. Rev. Res.},
  volume = {6},
  issue = {2},
  pages = {023061},
  numpages = {9},
  year = {2024},
  month = {Apr},
  publisher = {American Physical Society},
  doi = {10.1103/PhysRevResearch.6.023061},
  url = {https://link.aps.org/doi/10.1103/PhysRevResearch.6.023061}
}

@article{Legero_2004,
  title = {Quantum Beat of Two Single Photons},
  author = {Legero, Thomas and Wilk, Tatjana and Hennrich, Markus and Rempe, Gerhard and Kuhn, Axel},
  journal = {Phys. Rev. Lett.},
  volume = {93},
  issue = {7},
  pages = {070503},
  numpages = {4},
  year = {2004},
  month = {Aug},
  publisher = {American Physical Society},
  doi = {10.1103/PhysRevLett.93.070503},
  url = {https://link.aps.org/doi/10.1103/PhysRevLett.93.070503}
}
\end{document}